\documentclass[aps,prl,twocolumn,groupedaddress]{revtex4-1}

\usepackage[xdvi]{graphicx}

\begin{document}


\title{Crossover from negative to positive shear rate dependence in granular friction}


\author{Osamu Kuwano}
\email[]{kuwano@eri.u-tokyo.ac.jp}
\affiliation{Earthquake Research Institute, University of Tokyo, Tokyo 113-0032, Japan}

\author{Ryosuke Ando}
\affiliation{Geological Survey of Japan, AIST, Tsukuba 305-8567, Japan}

\author{Takahiro Hatano}
\affiliation{Earthquake Research Institute, University of Tokyo, Tokyo 113-0032, Japan}


\date{\today}

\begin{abstract}
We conduct an experiment on the frictional properties of granular matter over a wide range of shear rate that covers both the quasistatic and the inertial regimes.
We show that the friction coefficient exhibits negative shear-rate dependence in the quasistatic regime, whereas the shear-rate dependence is positive in the inertial regime.
This crossover from negative to positive shear-rate dependence occurs at a critical inertial number.
This is explained in terms of the competition between two physical processes, namely frictional healing and anelasticity.
We also find that the result does not depend on the shape of the grains and that the behavior in
the inertial regime is quantitatively the same as that in numerical simulations.
\end{abstract}


\maketitle

The rheological properties of granular matter are important for analyzing various phenomena in geosciences, such as landslides, debris flows, and crater formation.
These properties may also be relevant to friction in faults because a fault contains fine rock powders that have been ground up by past fault motions.
Among the rheological properties of granular matter, the shear-rate dependence of the friction coefficient determines the stability of deformation and therefore has been paid major attention.
During the last decade, considerable progresses have been made on this subject;
it is recognized that a nondimensional number plays a central role in describing the shear-rate dependence of the friction coefficient \cite{GDR2004,daCruz2005}.
This is referred to as the inertial number \cite{Savage1984}; $I=\dot\gamma\sqrt{m/Pd}$, where $\dot\gamma$ is the shear rate, $P$ is the pressure, $m$ is the mass of the grains, and $d$ is the diameter of the grains.
As $\sqrt{m/Pd}$ is the velocity relaxation time in granular matter \cite{Hatano2009}, the inertial number may be regarded as the index for the inertial effects inside the flow and therefore one can classify the flow regimes according to the value of the inertial number.
The flow is fast and collisional at larger inertial numbers ($I\ge 10^{-1}$), which is referred to as the inertial regime.
At lower inertial numbers ($I < 10^{-1}$), which is conventionally referred to as the quasistatic regime, the grains are closely packed and slowly sheared.
In the inertial regime, a constitutive law obtained in numerical simulation quantitatively explains the experimental observation on an inclined plane flow \cite{Jop2006}.
However, we have not reached the consensus about a constitutive law for the quasistatic regime \cite{Divoux2007,Behringer2008,Petri2008}.
Although some numerical studies \cite{daCruz2005,Hatano2007,Peyneau2008,Koval2009} are conducted down to $I\sim10^{-5}$ to address a constitutive law for the quasistatic regime, they are yet to be verified in experiments.
In addition, we are not aware of any constitutive laws that can lead to the exponential velocity profile that is ubiquitously observed in heap flows \cite{Lemieux2000,Komatsu2001}.

\begin{figure}
\includegraphics[width=18pc]{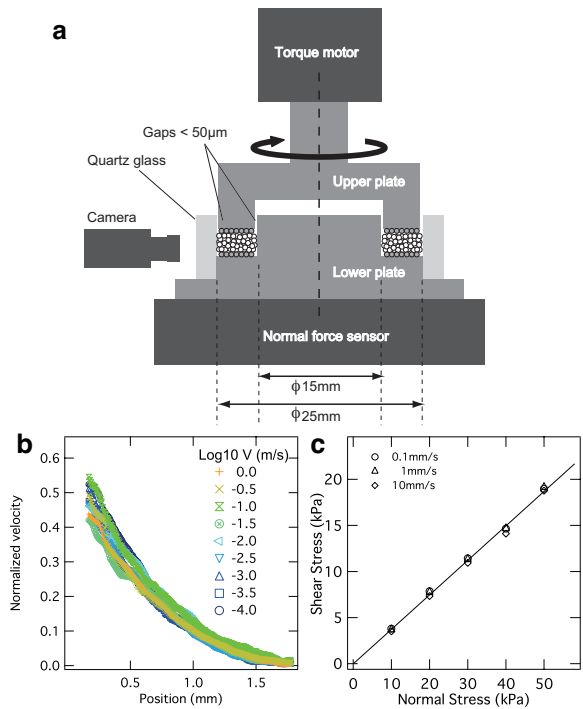}
\caption{
\label{fig:Apparatus}
(a) Schematic diagram of the apparatus.
A granular sample fills the annular channel, which is attached to a rheometer.
Shear is applied to the sample by rotating the upper plate, whereas the lower plate is fixed.
To avoid slipping on the surfaces, a monolayer of the glass beads was glued to both the upper and the lower plates.
(b) The velocity profiles normalized by the applied sliding velocities, $V$, ranging from $10^{-4}$ to $1$ m/s.
(c) The normal-stress dependence of the shear stress.}
\end{figure}

To clarify the nature of granular friction in the quasistatic regime and its connection to the constitutive law in the inertial regime, one must conduct the rheological measurements over a wide range of shear rates that cover the both regimes.
To this end, we used a commercial rheometer (AR-2000ex, TA Instruments), with which we can control the sliding velocity over a wide range ($1\times10^{-5}-3 \,\mathrm{m/s}$).
The normal force is also controllable.
We chose a relatively low pressure ($20-50$ kPa) to exclude the effects of frictional heat 
\cite{note:heat}.
In addition, the temperature beneath the lower plate was set to $25^\circ$C and was kept constant with Peltier Plate.
To study the effects of the grain shape on the rheological properties, we prepared two kinds of grains: soda-lime glass beads (spherical with a mean diameter of $d=270\,\mathrm{\mu m}$) and chromite sand (angular shaped with a mean diameter of $d=286\,\mathrm{\mu m}$).
The grains were packed into the annular channel of the sample holder \cite{note:Ve} as shown in FIG. \ref{fig:Apparatus} a.
The cylindrical side wall, which was fixed to the lower plate, is made of transparent fused silica so that we could optically observe the internal structure of sheared granular matter \cite{note:gap}.
The images of each grain were captured by a high-speed video camera (Photron Fastcam APX RS).

To observe the internal velocity profile, we used particle tracking velocimetry to obtain the instantaneous velocity of each grain, which was then averaged out with respect to the flow direction.
The velocity profiles are shown in FIG. \ref{fig:Apparatus} b.
Note that the velocity profiles that are normalized by the sliding velocity $V$ collapsed to a single curve.
As the velocity profiles were somewhat nonlinear as observed in \cite{Aharonov2002}, we defined the effective shear rate $\dot\gamma=V/W_s$ using the effective flow width $W_s$, which was defined as the depth at which the normalized flow velocity decreased to $1/10$.
We estimated $W_s\simeq 5d$.
Throughout this study, the inertial number is defined in terms of the effective shear rate.

\begin{figure}
\includegraphics[width=15pc]{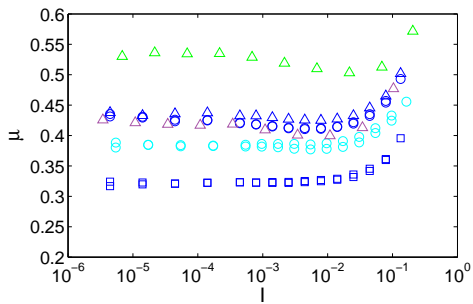}
\caption{ \label{fig:mu}
The friction coefficient plotted against the inertial number.
Symbols show the different normal stress (light blue circles: glass beads 20 kPa, 
blue circles/triangles/squares: glass beads 30 kPa, 
purple triangles: glass beads 50 kPa,
green triangles: chromite sand 30 kPa). }
\end{figure}

The shear stress $\sigma$ and the layer thickness $H$ were measured at each steady state of the sliding velocity $V$ under a constant normal stress $P$.
Fig. \ref{fig:Apparatus} c shows an example of the normal-stress dependence of the steady-state shear stress.
The intercept of the best-fit line is quite close to the origin and therefore the resistance from the side walls may be negligible.
We then define the friction coefficient $\mu$ as $\sigma/P$.
In Fig. \ref{fig:mu}, the friction coefficient is shown as a function of the inertial number.

At lower inertial numbers ($I\le10^{-2}$), the friction coefficient is weakly dependent of the sliding velocity.
Note that the dependence is negative and logarithmic.
The extent of weakening is characterized by the slope $\partial\mu/\partial\ln I$.
For example, the slope is approximately $-0.003$ for spherical glass beads at $30$ kPa.
The negative dependence on the sliding velocity is indeed well known in earthquake physics \cite{Marone1998} and in tribology \cite{Heslot1994} and the extent of weakening in the present study is comparable to those typically observed for glass or rocks, although the slope may be sensitive to the experimental details, such as humidity.

Although the negative slope is the common feature of the present data, the absolute value of the friction coefficient differs significantly from sample to sample. An apparent ingredient that affects the absolute value is the grain shape \cite{Mair2002,Anthony2005}. The friction coefficient of the angular grains ($\mu \simeq 0.55$) is significantly higher than that of the spherical grains ($\mu \simeq 0.4$). It is also  noteworthy that the spherical glass beads may exhibit anomalously low friction ($\mu\simeq 0.3$) despite the same experimental conditions (e.g., the blue squares in Fig. 2a). This effect may be due to the structural ordering induced by the shear \cite{Grebenkov2008}, although that hypothesis has not been verified.

\begin{figure}
\includegraphics[width=21pc]{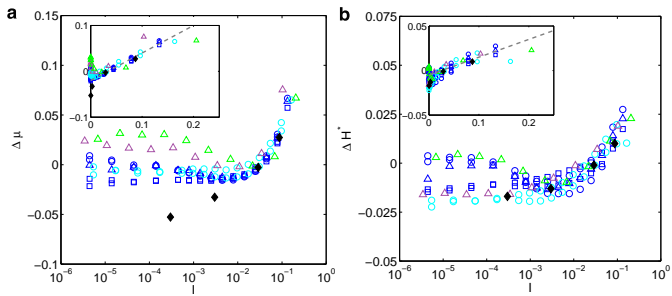}
\caption{\label{fig:UCurve}
(a) The increase of the friction coefficient, $\Delta \mu(I) (= \mu(I) - \mu(I_c))$, where $I_c(=0.032)$ is the crossover inertial number.
The inset is the linear plot.
(b) The normalized thickness change $\Delta H^*(I)$. Symbols are the same as those in the previous figure. The inset is the linear plot. Results of the DEM simulation with an inter-granular friction coefficient of 0.6 and a restitution coefficient of 0.8 are also shown (black diamonds).}
\end{figure}

\begin{figure}
\includegraphics[width=10pc]{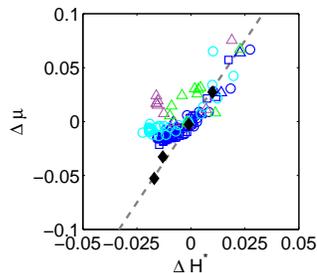}
\caption{
\label{fig:MGAP}
Friction coefficient difference $\mathsf{\Delta \mu}$ versus normalized thickness change $\mathsf{\Delta H^*}$.
The dashed straight line indicates a slope of 3.
Symbols are the same as those in the previous figures.}
\end{figure}

At higher inertial numbers ($I\ge10^{-2}$), the friction coefficient substantially increases and 
the data indicate the characteristic inertial number above which the friction coefficient increases.
We refer to this as the crossover inertial number, and it is denoted by $I_c$.
Although $I_c$ for each sample varies slightly, its value is apparently on the order of $10^{-2}$.
Thus, for simplicity, we chose $I_{\mathrm{c}} = 0.032$ assuming that $I_c$ is common to all of the data.
We then defined the amount of strengthening as $\Delta\mu(I) = \mu(I) - \mu(I_c)$ and found that $\Delta\mu(I)$ collapses in the high $I$ regime ($I\ge I_{\rm c}$) (as shown in Fig. \ref{fig:UCurve}a).
This collapse obeys
\begin{equation}
\label{strengthening}
\Delta \mu(I) \propto  c_1 I,
\end{equation}
with $c_1 \simeq 0.6$.
This strengthening behavior for $I\ge I_c$ is accompanied by dilation.
To compare the data with different layer thicknesses, it is convenient to define the nondimensional dilation as $\Delta H^* = (H(I)-H(I_c))/W_s$.
As shown in Fig. \ref{fig:UCurve} b, $\Delta H^*$ also collapses in the high $I$ regime.
\begin{equation}
\label{dilation}
\Delta H^* \propto  c_2 I,
\end{equation}
where $c_2 \simeq 0.2$.
From Eqs. (\ref{strengthening}) and (\ref{dilation}), it follows that 
\begin{equation}
\label{strengthening-dilation}
\Delta \mu = c_3 \Delta H^*,
\end{equation}
where $c_3=c_1/c_2 \simeq 3$.
Figure \ref{fig:MGAP} shows that Eq. (\ref{strengthening-dilation}) explains the experimental data well.
In addition, we noticed that Eqs. (\ref{strengthening}), (\ref{dilation}), and (\ref{strengthening-dilation}) quantitatively explain the inclined-plane flow data \cite{Forterre2008}.
Note that the proportional coefficients ($c_1$, $c_2$, and $c_3$) are common to these different experimental data.
We thus conclude that these relations are independent of the details of granular matter such as the grain shape and the coefficient of restitution.

Relations (\ref{strengthening}), (\ref{dilation}), and (\ref{strengthening-dilation}), which hold for the high $I$ regime, are quantitatively the same as those seen in numerical simulations \cite{GDR2004,daCruz2005,Jop2006,Hatano2007,Peyneau2008,Koval2009}.
The data obtained from the numerical simulations \cite{Hatano2007} are also shown in FIGS. \ref{fig:UCurve} and \ref{fig:MGAP}, which show that the simulations quantitatively agree with the experiments in the high $I$ regime.
It is believed that the velocity strengthening seen in numerical simulations is caused by the anelasticity of the grains, which is usually modeled by a "dashpot" \cite{Hatano2011}.
The quantitative agreement between the simulation and the experiment in the high $I$ regime implies that anelasticity plays an essential role in the velocity-strengthening behavior seen in the experiments.

In contrast, in the low $I$ regime, the numerical simulation no longer agrees with the experiments (as shown in FIG. \ref{fig:MGAP}.)
This indicates that a physical process comes into play that is not modelled in the simulations.
It is known that the negative shear-rate dependence is due to the healing of the contact between grains \cite{Brechet1994,Bocquet1998}.
Note that this process is not implemented in the numerical simulations, and this must be the reason for the disagreement.
Thus, the frictional healing process must be properly incorporated into numerical models that involve the low $I$ regime.

Taking the above points into account, we propose a constitutive law that holds for both the high and the low $I$ regimes.
\begin{equation}
\label{constitutivelaw}
\mu = c_1 I - \alpha \ln \dot\gamma\tau + \mu_0, 
\end{equation}
where $\tau$ is the characteristic time for frictional healing.
The first term on the right-hand side originates from the dissipation due to anelasticity, and the remaining terms are due to the dissipation caused by intergranular friction.
The linear combination of these two effects may be justified if the intergranular friction force and the viscous force (due to anelasticity) are orthogonal.
From Eq. (\ref{constitutivelaw}), the crossover inertial number $I_c$ is given  by
\begin{equation}
\label{Ic}
I_c = \frac{\alpha}{c_1}.
\end{equation}
Constitutive law (\ref{constitutivelaw}) and the critical inertial number (\ref{Ic}) explain the experimental data well.
We conclude that the switch of the dominant physical processes at $I\simeq I_c$ leads to the crossover from negative to positive shear-rate dependence of the friction coefficient;
the anelasticity is dominant over the healing for $I\ge I_c$, whereas the frictional healing is essential for $I\le I_c$.
As the crossover is located at $I \simeq 10^{-2}$, it is not surprising that it has not been identified in previous experiments on either inclined-plane flow ($I\ge 10^{-2}$) \cite{Forterre2008} or simulated fault gouge ($I\ll 10^{-2}$) \cite{Marone1998,Blanpied1987,Kilgore1993,Dieterich1994,Nakatani2001}.
Constitutive law (\ref{constitutivelaw}) and the resultant crossover constitute the main conclusion of this paper.
In the following, we discuss four important points that are peripherally related to this main conclusion.

First, we remark that the crossover cannot be explained within the framework of a phenomenological constitutive law known in earthquake physics \cite{Marone1998}.
It is known that a similar crossover may occur if the shear rate exceeds the characteristic rate of frictional healing, $1/\tau$ \cite{Blanpied1987,Kilgore1993}; in this case, the steady-state friction coefficient is written as $\mu = \mu_0 + \beta \ln V$ with $\beta > 0$.
If we were to adopt this relation to explain the present data in the high $I$ regime, the parameter $\beta$ would be much larger ($0.06$) than the typical values ($0.005-0.015$) that are reported for rocks and soda-lime glass \cite{Dieterich1994,Nakatani2001}.

Second, we remark that Eq. (\ref{strengthening-dilation}) is different from the well-known relation $\mu = \mu_* + dH / dx$, where $\mu_*$ is a constant and $x$ is the shear displacement of a boundary.
The latter states that the increase in the friction coefficient is due to the additional work required for dilation \cite{Marone1990}.
However, it does not state anything about the steady-state friction coefficient because $dH / dx=0$ at steady states (except for fluctuations), whereas Eq. (\ref{strengthening-dilation}) involves the difference between two steady states.

The third point we wish to discuss is with regard to the mechano-chemical reactions caused by frictional heat, which generally reduce friction to a considerable degree \cite{Tullis2007}.
We believe that anelasticity-induced strengthening is also relevant to such systems because mechano-chemical reactions affect the nature of intergranular friction, which is orthogonal to and independent of anelasticity.
A more detailed study on this issue is in progress.

The fourth point of interest is the application of the present result to other flow geometries.
Among them, here we derive the exponential velocity profile that is ubiquitously seen in heap flow experiments \cite{Lemieux2000,Komatsu2001}.
We begin with the force balance equation in the flow direction.
\begin{equation}
\label{forcebalance}
\frac{d\sigma}{d h}= \rho g \sin\theta,
\end{equation}
where $\rho$ is the mass density, $g$ is the gravitational acceleration, $\theta$ is the angle between the flow direction and the horizontal plane, $\sigma$ is the shear stress, and $h$ is the depth.
The existence of the body force in the flow direction distinguishes the heap flow from the annular channel flow.
In deriving the exponential velocity profile, it is essential to assume Janssen's law; i.e., the pressure within the heap, $P$, is independent of the depth due to the frictional force of a container.
Then Eq. (\ref{forcebalance}) together with Janssen's law leads to 
\begin{equation}
\label{forcebalance2}
\frac{d\dot\gamma}{dh}\frac{d\mu}{d\dot\gamma} = \frac{\rho g \sin\theta}{P}.
\end{equation}
Inserting Eq. (\ref{constitutivelaw}) into Eq. (\ref{forcebalance2}), we obtain $\dot\gamma\simeq\dot\gamma_0e^{-Ah}$, where $A=\rho g\sin\theta/\alpha P$.
If Janssen's law does not hold, the friction coefficient is $\tan\theta$ and independent of the depth. 
Then Eq. (\ref{constitutivelaw}) leads to a linear velocity profile, which is also observed in some experiments \cite{Ancey2001,Silbert2003}.

\appendix

\begin{acknowledgments}
This research was partially supported by JSPS KAKENHI (21840022) and Fukada Geological Institute.
\end{acknowledgments}


\begin{thebibliography}{28}
\bibitem{note:Ve}
Because sliding velocity varies with radial position on the upper plate, we introduced an equivalent sliding velocity $V_e=(\dot\theta/3)\times(D_2^3-D_1^3)/(D_2^2-D_1^2)$, where $\dot\theta$ is the angular velocity, and $D_1$ and $D_2$ are the inner and the outer diameters, respectively.
With this definition and the assumption of the uniform shear stress $\tau$ over the sliding area $S$, the rate of the total frictional work equals $\tau V_eS$.
The equivalent sliding velocity is referred to simply as the sliding velocity $V$ throughout this study.
\bibitem{note:gap}
The gaps between the upper plate and the side wall are smaller than $50$ $\mu{\rm m}$, which is four times smaller than the grain diameter.
\end{thebibliography}

\end{document}